# Theoretical study on hot-carrier generation, transport and injection in TiN/TiO$_2$ junction

Tingting Liu[1,2#], QianJun Wang[3#], Cheng Zhang[1,2*], Xiaofeng Li[1,2*], Jun Hu[3,4*]

[1] *School of Optoelectronic Science and Engineering & Collaborative Innovation Center of Suzhou Nano Science and Technology, Soochow University, Suzhou 215006, China;*

[2] *Key Lab of Advanced Optical Manufacturing Technologies of Jiangsu Province & Key Lab of Modern Optical Technologies of Education Ministry of China, Soochow University, Suzhou 215006, China;*

[3] *School of Physical Science and Technology, Soochow University, Suzhou 215006, China.*

[4] *School of Physical Science and Technology, Ningbo University, Ningbo 315211, China.*

**Abstract:** Promoting performance of generation, transport and injection of hot carriers in metal/semiconductor junctions is critical for harvesting energy of hot carriers. However, the injection efficiency of hot carriers in the commonly used noble metals such as Au is extremely low, which hinders the hot carrier-based applications. Here, we predict that the plasmonic material TiN might be promising for generating, transporting and injecting hot carriers, based on first-principles calculations and Monte Carlo simulations. We found that the hot-carrier generation rate in TiN is significantly large compared to Au. The transport property of hot carriers in TiN is excellent, due to the long lifetime and mean free path. The TiN/TiO$_2$ junction possesses very low Schottky barrier, which results in much larger injection efficiency than that in the Au/TiO$_2$ junction. Furthermore, it is revealed that the optimal injection efficiency could be achieved in a core/shell cylindrical TiN/TiO$_2$ junction. Our findings provide an in-depth understanding of hot-carrier generation, transport and injection, and hence are helpful for the development of hot carrier-based devices.

# 1. Introduction

Surface plasmons in a metal/semiconductor (M/S) junction excited by light irradiation have sparked great interest in the communities of condensed matter physics and materials science recently, because of the promising applications for solar energy harvest, photo-electric detection and photocatalysis [1-3]. Decay of surface plasmons in the metal of an M/S junction may generate hot carriers which could be injected into the semiconductor by overcoming a Schottky barrier at the M/S interface [4]. Consequently, the photon energy requisite to obtain free carriers in the semiconductor of an M/S junction through the hot-carrier injection could be much lower than that of direct interband transitions from valence bands to conduction bands. The noble metals especially Au are most widely used in hot-carrier devices, due to their excellent environmental stability, good electrical conductivity, high plasmon enhanced hot-carrier generation rate, and long lifetime and mean free path of hot carriers [5-8]. However, Au is not an ideal plasmonic metal for practical hot-carrier devices. Firstly, it is an expensive and scarce resource on the earth. Secondly, the density of the light-excited hot carriers near the Fermi level is very low, because of the specific electronic structures of Au. Thirdly, the work function of Au is high, which usually results in large Schottky barrier and hence makes the injection efficiency of hot carriers in Au-based M/S junctions extremely low [7]. Accordingly, exploring better plasmonic materials is of high importance and interest for the development of hot carrier-based devices.

Revealing the underlying physical mechanisms of the generation, transport and injection properties of hot carriers in M/S junctions is the prerequisite for designing hot-carrier devices [9-11]. These properties are closely associated with the electronic structures of both metal and semiconductor. The light irradiation excites electrons from occupied states to unoccupied states, including interband and intraband transitions. Hence, the energy distribution of hot carriers is determined by the band structure of the metal and the photon energy of the incident light [12, 13]. The transport property is related to the lifetimes and mean free paths of hot carriers, so it is significantly affected by the scatterings from electrons, phonons, defects and impurities. Generally, defects and impurities might be avoided by controlling the quality of M/S junctions, thus they are not the crucial factors for the transport of hot carriers [14]. However, the scatterings from electrons and phonons are inevitable, and they play important roles in the transport procedure. The hot-carrier injection efficiency is determined by the Schottky barrier at the M/S interface which stems from the band alignment between the metal and semiconductor [14, 15]. Clearly, revealing the corresponding underlying physical mechanisms rely on accurate calculations of the electronic properties especially the

electron- and phonon-related scatterings. However, these calculations are still difficult and the systematic research on the whole process of generation, transport and injection of hot carriers in M/S junctions is still lacking [7, 12, 13].

In this paper, we provide a standard procedure to explore the generation, transport and injection performances of hot carriers in plasmonic material-based M/S junctions. We chose $TiO_2$ as the semiconductor layer, because it possesses good electronic properties and has been widely used in hot carrier-based devices [16-19]. For the plasmonic material, the family of the transition metal nitrides may be good candidates to replace the noble metals [20-25]. In particular, TiN has drawn extensive attention in recent years, because it exhibits excellent capability in photoelectric conversion and detection [26-28]. Especially, it is cheap compared to the noble metals. The generation, transport and injection of hot carriers in the $Au/TiO_2$ and $TiN/TiO_2$ junctions were investigated through first-principles calculations and Monte Carlo simulations [29-31]. We found that TiN is indeed a better plasmonic material than Au. TiN possesses higher concentration of low-energy hot carriers than Au, and the lifetime and mean free path of the hot carriers are comparable to those in Au. Large hot-carrier injection efficiency can be achieved in the core/shell cylindrical $TiN/TiO_2$ junction. Accordingly, the $TiN/TiO_2$ junction might be promising for practical hot-carrier applications.

## 2. Calculation Methods

We used the Vienna Ab initio Simulation Package (VASP) [32, 33] in the framework of the density functional theory (DFT) to calculate the atomic structures and electronic properties. The electron-ion interactions were described with the projector augmented wave (PAW) approach [34]. The exchange-correlation potential was considered at the level of the generalized gradient approximation (GGA) with Perdew-Burke-Ernzerhof (PBE) functional for TiN, and Perdew-Burke-Ernzerhof revised for solids (PBEsol) [35] functional for Au. According previous report, the spin-orbit coupling (SOC) and a rotationally invariant DFT+U correction [36] with U = 2.4 eV is necessary for Au [12, 13]. A kinetic energy cutoff was set to 500 eV and a 10 × 10 × 10 Monkhorst-Pack mesh in the Brillouin Zones was employed. The calculated electronic structures of Au and TiN are consistent with previous calculations and experiments [37, 38]. The crystal structures were fully relaxed until the residual forces on each atom are less than 0.01 eV/Å. The optimized lattice constants of Au and TiN are 4.06 Å and 4.23 Å, respectively, consistent with the corresponding experimental values 4.08 Å [39] and 4.24 Å [40]. The electron- and phonon-related scatterings were estimated by using the open-source code

JDFTx [41] with full-relativistic norm-conserving pseudopotentials [42]. The plane-wave cutoff and Fermi-Dirac smearing were set to 30 and 0.01 Hartree, respectively. The Brillouin zone was sampled with a 20 × 20 × 20 Monkhorst-Pack mesh. The matrix elements of the electron-phonon coupling were calculated with the maximally-localized Wannier functions in a 2 × 2 × 2 supercell [43]. The k-point mesh for the Wannier function interpolation was adopted as 24 × 24 × 24.

## 3. Results and discussion

### 3.1 Electronic properties of bulk Au and TiN

The bulk Au adopts the face-centered cubic (FCC) crystal structure, while TiN crystallizes the rock-salt structure which is also based on the FCC structure, as shown in Figure 1(a) and 1(b). Apparently, each Ti atom locates at the center of the octahedron comprised of the six nearest neighboring N atoms, and vice versa for each N atom. Consequently, the five-fold degenerate Ti-3d orbital splits into two groups: triplet $t_{2g}$ ($d_{xy}$, $d_{xz}$, and $d_{yz}$) and doublet $e_g$ ($d_{z^2}$ and $d_{x^2-y^2}$) orbitals, due to the octahedral crystal field. Then, the $t_{2g}$ and $e_g$ orbitals separately hybridize with the N-2p orbital, forming bonding and antibonding orbitals, as sketched in Figure 1(d). These hybridized orbitals finally evolve into the energy bands due to the periodic potential in crystal TiN.

The band structure of Au with the path denoted in Figure 1(c) is displayed in Figure 2(a). We can see that there are many localized bands in the energy range from -7 eV to -1 eV, while only a few bands cross the Fermi level ($E_F$). To identify the orbital features of these bands, we plotted the orbital-resolved band structures in Figure 2(b) and 2(c). It can be seen that the localized bands bellow $E_F$ are dominated by the Au-5d orbital. Interestingly, the Au-6p orbital has large contribution to the bands crossing $E_F$, which originates from the charge transfer between different atomic orbitals due to the crystal field and hybridization between neighboring Au atoms.

Figure 2(d-f) shows the element- and orbital-resolved band structures of TiN. It can be seen that the bands can be separated into two parts. The upper bands from -1 eV to 6 eV are dominated by the Ti-3d orbital, while the lower bands bellow -2 eV are mainly contributed from the N-2p orbitals. Meanwhile, the energy levels at the Γ point remain completely as atomic orbitals. The three energy levels in the oval in Figure 2(d) at -2.03 eV, -0.63 eV, and 0.57 eV are the N-2p, Ti-$t_{2g}$, and Ti-$e_g$ orbitals, respectively. Away from the Γ point, the bands reflect strong hybridization between the orbitals of Ti and N, because of the short Ti-N bond length (2.12 Å). Moreover, the upper bands can be assigned to the bonding states of the *p*-t$_{2g}$ and *p*-e$_g$ hybridizations, compared to the energy diagram in Figure 1(d). The hybridizations for the lower bands are more complicated, since the N-2p orbital hybridizes with all the orbitals (*s*, *p* and *d*) of Ti.

The band alignment of the metal and semiconductor of an M/S junction, which determines the Schottky barrier at the M/S interface, is important for the transport of hot carriers. Based on the band structures of TiO$_2$ and TiN, we found that the Schottky barrier at the TiN/TiO$_2$ interface is only about 0.1 eV [44], much smaller than that at the Au/TiO$_2$ interface (~1 eV), in good agreement with previous report [45]. Accordingly, the band gap of TiO$_2$ covers the energy range from -3.1 to 0.1 eV with respect to $E_F$ of TiN, while from -2.1 to 1.1eV for Au.

The total and projected density of states (DOS) of Au and TiN are plotted in Figure 3(a) and 3(b). Clearly, the total DOS near $E_F$ in TiN is much larger than that in Au, because there are more bands crossing $E_F$ in TiN than in Au. Therefore, the concentration of low-energy free carriers in TiN is much higher than that in Au, and the free carriers mainly characterize the Ti-3d electrons. Moreover, we can see that the peak of DOS within 0 ~ 3 eV features the Ti-t$_{2g}$ orbital, while that within 3 ~ 6 eV belongs to the Ti-e$_g$ orbital, according to the energy diagram in Figure 1(d).

Fermi surface (FS) of metals is another important electronic property which affects the electronic transport in metals. The insets in Figure 3(a) and 3(b) display the FSs of Au and TiN, respectively. Clearly, most part of the FS of Au is smooth and far away from the boundary of Brillouin zone. The FS of Au touches the boundary of Brillouin zone within a small region around the "L" point and its equivalents, which contributes to the electron transport. In contrast, the FS of TiN is much more complex compared to that of Au, especially near the boundary of Brillouin zone. Consequently, the anisotropy of electronic transport in TiN is more significant than in Au.

## 3.2 Hot-carrier generation

The hot-carrier generation rate is proportional to the relative probabilities of the indirect intraband and direct interband transitions. The indirect transition is associated to the DOS [$\rho(E)$] and Fermi-Dirac distraction function [$f(E)$] [46], and the corresponding relative probability of indirect transition is expressed as [47]:

$$D(E) = \frac{\rho(E-h\upsilon)f(E-h\upsilon)\rho(E)[1-f(E)]}{\int \rho(E-h\upsilon)f(E-h\upsilon)\rho(E)[1-f(E)]dE}, \tag{1}$$

where $E$ is the energy of the excited electron and $h\nu$ is the incident photon energy. The direct transition is usually described by the imaginary part of the dielectric function ($\bar{\epsilon}_{direct}$). Based on the perturbation theory of quantum mechanics for the electric dipole transition, the imaginary part of $\bar{\epsilon}_{direct}$ can be calculated with the following expression [13]:

$$\lambda^* \cdot \text{Im}\bar{\epsilon}_{direct}(\omega) \cdot \lambda = \frac{4\pi^2 e^2}{m_e^2 \omega^2} \int_{BZ} \frac{d\boldsymbol{k}}{2\pi^3} \sum_{n'n} (f_{\boldsymbol{k}n} - f_{\boldsymbol{k}n'}) \delta(\varepsilon_{\boldsymbol{k}n'} - \varepsilon_{\boldsymbol{k}n} - \hbar\omega) |\lambda \cdot \langle \boldsymbol{P} \rangle_{n'n}^{\boldsymbol{k}}|^2,$$

(2)

where $\lambda$ and $\omega$ are the polarization vector and angular frequency of incident light, respectively; $\varepsilon_{\boldsymbol{k}n}$ and $f_{\boldsymbol{k}n}$ are the eigenvalue and electron occupancy of the state with wave vector $\boldsymbol{k}$ and band index $n$; $\langle \boldsymbol{P} \rangle_{n'n}^{\boldsymbol{k}}$ stands for the matrix element of the momentum operator $\boldsymbol{P}$ which can be calculated via $\boldsymbol{P} \equiv \frac{m_e}{i\hbar}[\boldsymbol{r}, \boldsymbol{H}]$. We used the WANNIER90 code [43, 48-49] to calculate the maximum local Wannier functions that were then used to calculate the matrix elements of $\boldsymbol{r}$ and $\boldsymbol{H}$. The Brillouin zone for the integration was sampled by randomly selected k points up to $5 \times 10^6$.

The curves of $\bar{\epsilon}_{direct}$ as a function of $h\nu$ for Au and TiN are shown in Figure 3(c). When $h\nu < 1.5$ eV, $\bar{\epsilon}_{direct}$ of TiN is much larger than that of Au, while the values of $\bar{\epsilon}_{direct}$ for the two cases are comparable in the rest energy range. For Au, there are three peaks centered around 1.7, 2.5 and 3.3 eV, as marked by "A", "B" and "C" in Figure 3(c). The peak "A" corresponds to the threshold of electric dipole transition in Au, in good agreement with the experimental measurement (1.6 ~ 1.8 eV) [13]. According to the selection rules of the electric dipole transition, these peaks are generated mainly by the transition from the occupied 5d orbital to unoccupied 6p orbital, as indicated in Figure 2(a) and highlighted in the inset in Figure 3(a). For TiN, the electric dipole transitions mainly occur between the Ti-3d and N-2p orbitals. Since the DOS near $E_F$ in TiN is much larger than that in Au, it is expected that higher density of hot carriers could be produced in TiN.

We set a series of $h\nu$ to estimate the relative probabilities of hot-carrier generation at different energies from indirect and direct transitions in Au and TiN, as plotted in Figure 4. Generally speaking, larger $h\nu$ results in larger energy of hot carriers, as the peaks of relative probabilities shift away from $E_F$ with increasing $h\nu$. For Au, hot holes (bellow $E_F$) and hot electrons (above $E_F$) can be excited only through the indirect intraband transition when $h\nu$ is smaller than 1.2 eV [Figure 4(a)]. This is because the starting point of the first peak of the direct interband transition is around 1.2 eV, as can be seen in Figure 3(a). Once the incident photon energy is larger than 1.2 eV, both indirect and direct transitions produce hot holes and hot electrons. However, the relative probability from the indirect transition is much smaller than that from the direct transition. Furthermore, when $h\nu$ is smaller than 2 eV, the energies of both hot electrons and hot holes except a small part of hot electrons from the indirect transition fall inside the band gap of $TiO_2$, so the number of effective hot carriers is ignorable. Only when $h\nu$ is larger than 2.6 eV, there is sizable number of effective hot holes with energy centered around -2 eV from the direct transition [Figure 4(b)]. Clearly, this requirement is disadvantageous for hot-carrier applications which is mostly in the range from the near-infrared light to the low-energy visible light.

Figure 4(c) and 4(d) present the relative probabilities of hot carriers with different energies from indirect and direct transitions in TiN. Since the band gap of $TiO_2$ covers the energy range from -3.1 to 0.1 eV, the number of hot holes is insignificant for all considered incident photon energies. On the contrary, almost all hot electrons may turn into effective hot carriers, because their energies fall into the conduction band of $TiO_2$. In addition, hot carriers can be generated by both indirect and direct transitions even at low-energy incident light, which can be attributed to the large DOS near $E_F$. The overall relative probabilities from the indirect and direct transitions are comparable with $h\nu$ smaller than 2.0 eV, much different from the feature in Au. When $h\nu$ reaches 2.0 eV, the relative probability from the direct transition exceeds that from the indirect transition, and the energy of hot holes becomes more centralized as indicated by the narrow and sharp peaks in Figure 4(d). It is possible to identify the contributions of the direct transitions to these peaks, combined with the band structures in Figure 2. For example, when $h\nu$ is 3 eV, the hot electrons with energy around 0.75 eV mainly originate from the transition from the Ti-3d orbital to the N-2p orbital near the Γ point, while the hot electrons with energies of 2, 2.5 and 3eV mainly stem from the transitions around the W, K, Γ points, respectively. On the whole, the generation of hot carriers in TiN is much more efficient than in Au, especially in the near-infrared light and low-energy visible light, which benefits the solar energy harvest with hot-carrier devices.

## 3.3 Hot-carrier transport and injection

The hot carriers excited in the metal of an M/S junction travel in the metal before they get to the M/S interface, then they might be injected into the semiconductor if their energy is large enough to overcome the Schottky barrier at the M/S interface. The transport process is dominated by the lifetime and mean free path of hot carriers, so it is significantly affected by the electron-electron and electron-phonon scatterings (notated as $S_{ee}$ and $S_{ep}$, respectively) [50]. For the sake of practical applications, long lifetime and mean free path are desired. The scattering rates of $S_{ee}$ and $S_{ep}$ are related to the imaginary parts of the corresponding quasiparticle self-energies [13, 51]. The imaginary part of quasiparticle self-energy from $S_{ee}$ can be expressed as [13, 37]:

$$\text{Im}\,\Sigma_{kn}^{e-e} = \int_{BZ}\frac{d\boldsymbol{k}'}{(2\pi)^3}\sum_{n'}\sum_{\boldsymbol{GG}'}\tilde{\rho}_{\boldsymbol{k}'n',kn}(\boldsymbol{G})\tilde{\rho}^*_{\boldsymbol{k}'n',kn}(\boldsymbol{G}')$$

$$\times \frac{4\pi e^2}{\left|\boldsymbol{k}'-\boldsymbol{k}+\boldsymbol{G}\right|^2}\text{Im}\left[\epsilon^{-1}_{\boldsymbol{GG}'}(\boldsymbol{k}'-\boldsymbol{k},\varepsilon_{kn}-\varepsilon_{k'n'})\right].$$

(3)

Here, $\tilde{\rho}_{\boldsymbol{k}'n',kn}$ is the plane-wave expansion of the product density $\sum_{\sigma}u^{\sigma*}_{\boldsymbol{k}'n'}(\boldsymbol{r})u^{\sigma}_{\boldsymbol{k}'n'}(\boldsymbol{r})$ of Bloch functions with reciprocal lattice vectors $\boldsymbol{G}$; $\epsilon^{-1}_{\boldsymbol{GG}'}$ is the microscopic dielectric function in a plane-wave basis calculated within the random-phase approximation. For $S_{ep}$, the imaginary part of the quasiparticle self-energy can be calculated as [13, 37]:

$$\text{Im}\,\Sigma_{kn}^{e-ph} = \pi\int_{BZ}\frac{\Omega d\boldsymbol{k}'}{(2\pi)^3}\sum_{n'\,\alpha\pm}\left(n_{\boldsymbol{k}'-\boldsymbol{k},\alpha}+\frac{1}{2}\mp\frac{1}{2}\right)\times\delta(\varepsilon_{k'n'}-\varepsilon_{kn}$$

$$\mp\hbar\omega_{\boldsymbol{k}'-\boldsymbol{k},\alpha})\left|g^{\boldsymbol{k}'-\boldsymbol{k},\alpha}_{\boldsymbol{k}'n',kn}\right|^2.$$

(4)

Here, $\boldsymbol{q}=\boldsymbol{k}'-\boldsymbol{k}$ and $n_{q,\alpha}$ are the wave vector and particle number of phonons; $g^{\boldsymbol{k}'-\boldsymbol{k},\alpha}_{\boldsymbol{k}'n',kn}$ is the electron-phonon coupling matrix element with electronic states labeled by electronic wave vectors $\boldsymbol{k}$, $\boldsymbol{k}'$ and band indices $n$, $n'$; $\Omega$ is the volume of unit cell. The total scattering lifetime of hot carriers is $\tau_{kn}=\hbar/(2\text{Im}\Sigma_{kn})$, where $\text{Im}\Sigma_{kn}=\text{Im}\Sigma_{kn}^{e-e}+\text{Im}\Sigma_{kn}^{e-ph}$ [13]. The mean free path is $\lambda_{kn}=v_{kn}\tau_{kn}$, where $v_{kv}\equiv\frac{\partial\varepsilon_{kn}}{\partial\boldsymbol{k}}$ is the velocity of hot carriers and $\varepsilon_{kn}$ is the eigenvalue of the state with wave vector $\boldsymbol{k}$ and band index $n$.

The imaginary part of the quasiparticle self-energy from $S_{ee}$ and $S_{ep}$ as well as the corresponding lifetime and mean free path are plotted in Figure 5. It can be seen that the imaginary part of self-energy of the hot holes in Au is larger than that in TiN, while the situation of the hot electrons is opposite. This is because Au and TiN have larger DOS bellow and above $E_F$, respectively. In addition, $S_{ee}$ and $S_{ep}$ lead to comparable self-energies for the hot holes in the considered energy range (-5 eV ~ 0), but the former gives rise to significantly larger self-energy for the hot electrons with energy higher than 2 eV. Consequently, $S_{ee}$ and $S_{ep}$ contribute to similar lifetime and mean free path of the hot holes, while for the hot electrons the lifetime and mean free path from $S_{ep}$ are larger than those from $S_{ee}$. Particularly, the lifetime and mean free path of the low-energy hot carriers ($|E| \lesssim 1$ eV) from $S_{ee}$ are one- to three-order greater than the others. This is because the imaginary part of self-energy related to this phenomenon is very small, indicating the weak $S_{ee}$ in this energy range. In fact, $S_{ee}$ for electrons near $E_F$ is nominally proportional to $(\varepsilon - E_F)^2$ due to the phase space available for scattering [13, 14], so that it is negligible. Nevertheless, this feature does not benefit the transport of hole carriers in Au, since the energies of the effective hot holes and hot electrons in Au are larger than 1 eV due to the band alignment between Au and $TiO_2$, as shown in Figure 4. On the contrary, the requirement for the effective hot electrons in TiN is only about 0.1 eV, so the lifetime and mean free path of the low-energy hot electrons are mainly limited by $S_{ep}$. Nonetheless, the overall lifetime and mean free path of the effective low-energy hot electrons ($0.1 < E < 1$) in TiN are comparable to the effective low-energy hot electrons in Au ($1 < E < 2$).

Now, we can estimate the injection efficiency ($\eta$) of hot carriers by using the Monte Carlo simulation proposed by Blandre *et al* [31]. We chose three structures for the Au/$TiO_2$ and TiN/$TiO_2$ junctions: planar bilayer, planar sandwich trilayer, and core/shell cylinder, as depicted in Figure 6. The sizes of the Au and TiN layers range from 5 nm to 100 nm, and the hot carriers are supposed to be uniformly distributed in Au and TiN. Then the Monte Carlo approach is performed for a mass of electrons (100000 in our work) to calculate the probabilities of $S_{ee}$ and $S_{ep}$. Unlike the assumption of uniform energy distribution of hot carriers in Ref. [31], we used the calculated energy distribution in Figure 4. Given that most solar energy is carried by the infrared and visible lights [52, 53], and the energy of the effective hot holes is much larger than that of the effective hot electrons (Figure 4), we focus on $\eta$ of the hot electrons with incident photon energy $h\nu$ smaller than 2.5 eV.

As shown in Figure 6, the TiN/$TiO_2$ junction possesses much larger $\eta$ than Au/$TiO_2$. This can be understood as follows. First, the band alignment between TiN and $TiO_2$ is better than that between Au and $TiO_2$, so the Fermi level of TiN is very close to the conduction band minimum of $TiO_2$. Second, the Schottky barrier at the TiN/$TiO_2$ interface is only 0.1eV, much smaller than that of the Au/$TiO_2$ interface. Third, TiN has much more effective hot electrons in this energy range than Au as shown in Figure 4. Therefore, TiN might be a promising plasmonic material to replace Au in practical hot carrier-based devices.

It can be seen that $\eta$ depends on the incident photon energy $h\nu$. For the Au/TiO$_2$ junction, the amplitude of $\eta$ increases monotonically as $h\nu$ increases. For the TiN/TiO$_2$ junction, the value of $\eta$ increases rapidly when $h\nu$ increases from 0.1 eV to 0.6 eV and reaches the maximum value around 0.6 eV, for all considered cases. As $h\nu$ further increases, $\eta$ decreases monotonically, due to the increasing S$_{ee}$ and S$_{ep}$. On the other hand, the thickness of the TiN layer has negative effect on $\eta$. At the same $h\nu$, $\eta$ decreases as the thickness of TiN increases. In fact, this is evident because the lifetime and mean free path of the hot electrons are almost independent of the thickness. However, the thickness of the TiN layer cannot be too thin, because the number of hot electrons will be limited. Accordingly, the optimal thickness of the TiN layer should be smaller than but close to the mean free path of the hot electrons.

The structure of the junction also affects $\eta$, as seen in Figure 6. First of all, $\eta$ in the planar sandwich trilayer TiN/TiO$_2$ junction is nearly twice of that in the planar bilayer TiN/TiO$_2$ junction with the same thickness of TiN. Apparently, both TiO$_2$ layers in the former collect the hot electrons, if we assume that the hot electrons are distributed uniformly in TiN. The core/shell cylindrical TiN/TiO$_2$ junction can further enhance $\eta$, which can be understood as follows. In general, the distribution of the velocities of the hot electrons is almost isotropic in TiN, so that all hot electrons that reach the TiN/TiO$_2$ interface can be injected into the TiO$_2$ layer in the core/shell cylindrical TiN/TiO$_2$ junction. In contrast, the hot electrons with velocity parallel to the TiN/TiO$_2$ interface in the planar TiN/TiO$_2$ junctions will be lost. Therefore, the core/shell cylindrical junction is preferred in practical hot-carrier devices, and very large $\eta$ is expected to be achieved.

## 4. Conclusions

In summary, we carried out first-principles calculations and Monte Carlo simulations to study the performance of the generation, transport and injection of hot carriers in the Au/TiO$_2$ and TiN/TiO$_2$ junctions. We found that TiN has much higher concentration of hot carriers in low energy range than Au, because of the high density of states around the Fermi level in TiN. The lifetime and mean free path of the hot carriers in TiN are comparable to those in Au, which benefits the hot-carrier transport. Interestingly, almost all hot electrons in TiN can be injected into TiO$_2$ as hot carriers, because of the specific band alignment and low Schottky barrier of ~ 0.1 eV of the TiN/TiO$_2$ junction. Compared to the Au/TiO$_2$ junction, the average energy of hot carriers in TiN/TiO$_2$ is much lower and the injection efficiency is much larger. The injection efficiency depends on the thickness of TiN, the structure of the junction and the incident photon energy. The optimal thickness of TiN should be close to the mean free path of the hot electrons, and the maximum injection efficiency is achieved in the core/shell cylindrical TiN/TiO$_2$ junction with the incident photon energy around 0.6 eV.

**Author information**

#These authors contributed equally: Tingting Liu, Qianjun Wang

Corresponding authors

*Email: hujun2@nbu.edu.cn.

*Email: xfli@suda.edu.cn.

*Email: zhangc@suda.edu.cn.

**Acknowledgements**

We really appreciate the financial support from the National Natural Science Foundation of China (61875143, 61905170, 62075146, and 11574223), Natural Science Foundation of Jiangsu Province (BK20180042, BK20181169 and BK20190816), Natural Science Foundation of the Jiangsu Higher Education Institutions of China (17KJA480004), Priority Academic Program Development (PAPD) of Jiangsu Higher Education Institution, and the start-up funding of Ningbo University (422111593).

**Conflict of interest**

The authors have no conflicts to disclose.

**Figures**

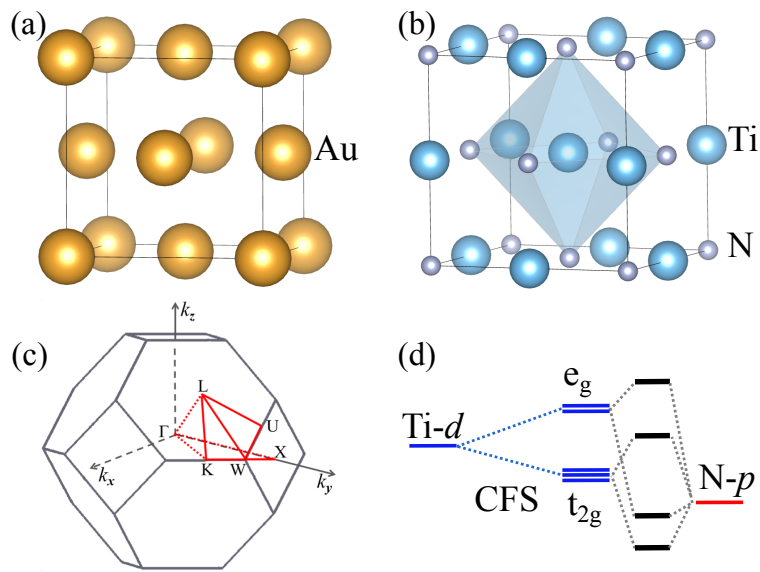

Figure 1. (a) and (b) Atomic structures of Au and TiN. (c) The Brillouin zone. The red lines denote the path for the band structure. (d) Schematic energy diagram of the hybridization between the Ti-3d and N-2p orbitals due to the crystal field splitting (CFS) in the octahedron symmetry as indicated in (b).

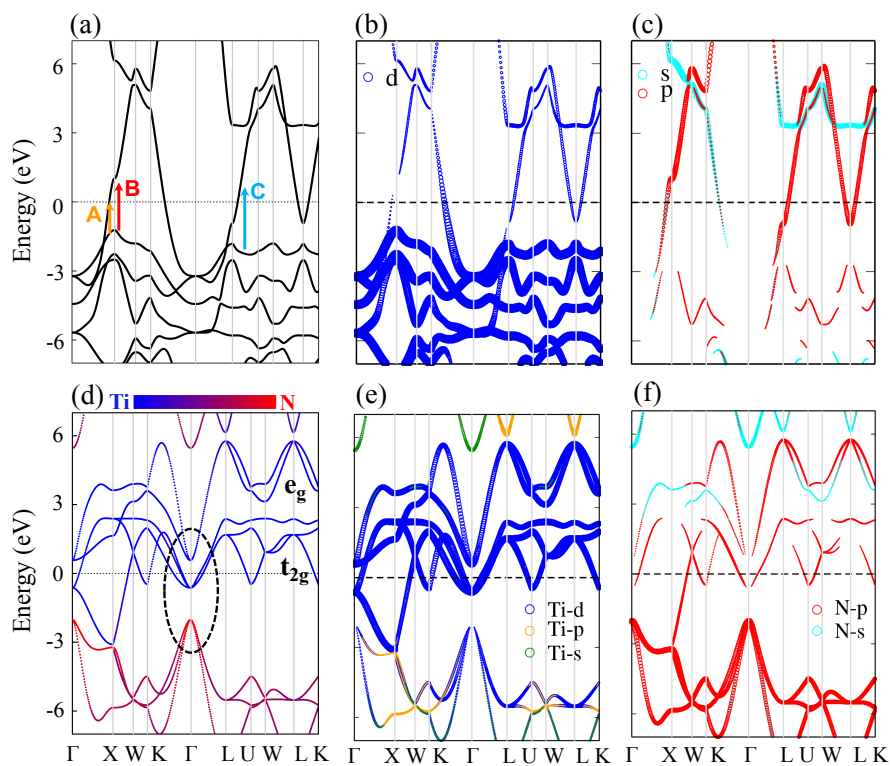

Figure 2. Band structures of Au (upper panels) and TiN (lower panels). The different colors and sizes of circles in (b), (c), (e) and (f) indicate the weights of different atomic orbitals. The color bar in (d) indicates the weights of Ti and N atoms.

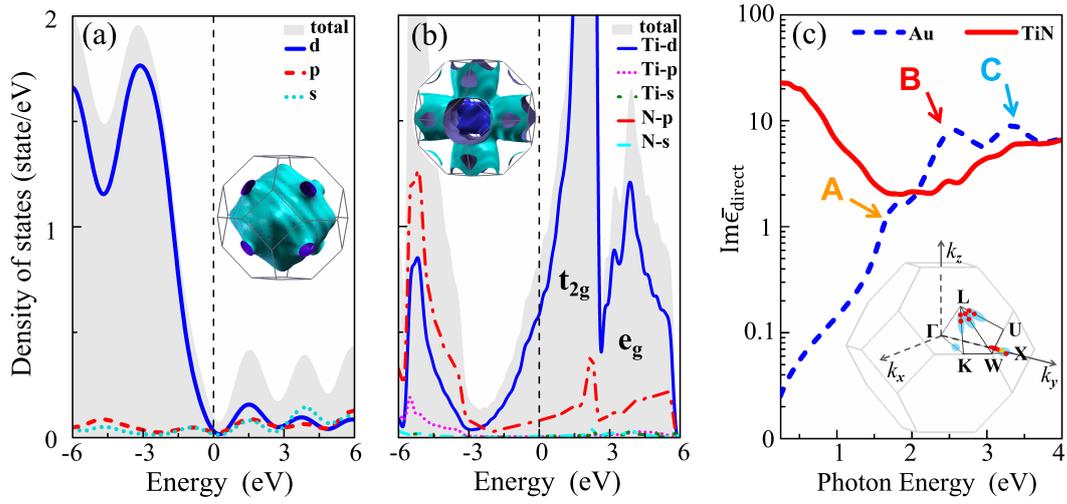

Figure 3. (a) and (b) Projected densities of states of Au and TiN, respectively. The insets are the corresponding Fermi surfaces. (c) Imaginary part of dielectric function for the direct electric dipole transition. The inset highlights the areas where the peaks "A", "B" and "C" mostly originate from.

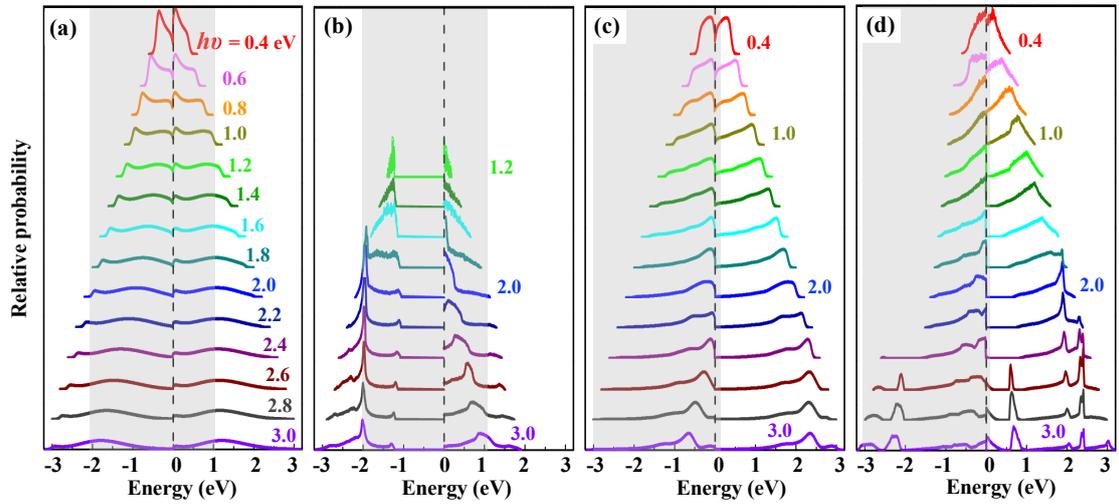

Figure 4. Relative probability of hot-carrier generation as a function of hot-carrier energy under different incident photon energy $h\nu$. (a/c) and (b/d) From indirect and direct transitions in Au/Ti, respectively. The Fermi level is set to zero energy. The shadow area denotes the band gap of $TiO_2$.

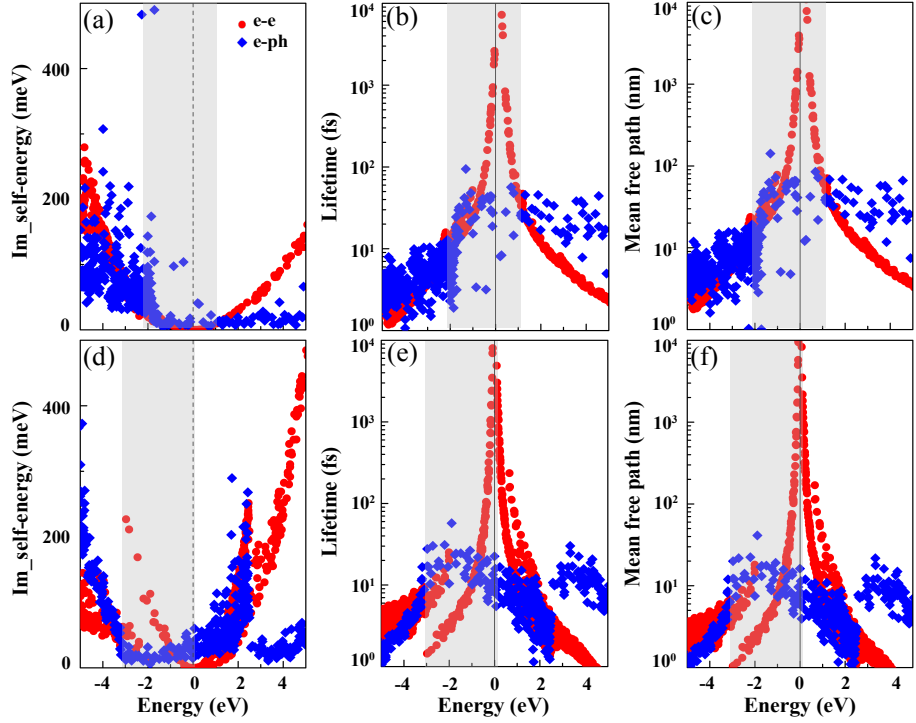

Figure 5. Imaginary part of quasiparticle self-energy, lifetime and mean free path of hot carriers from electron-electron and electron-phonon scatterings in Au (upper panels) and (b) TiN (lower panels).

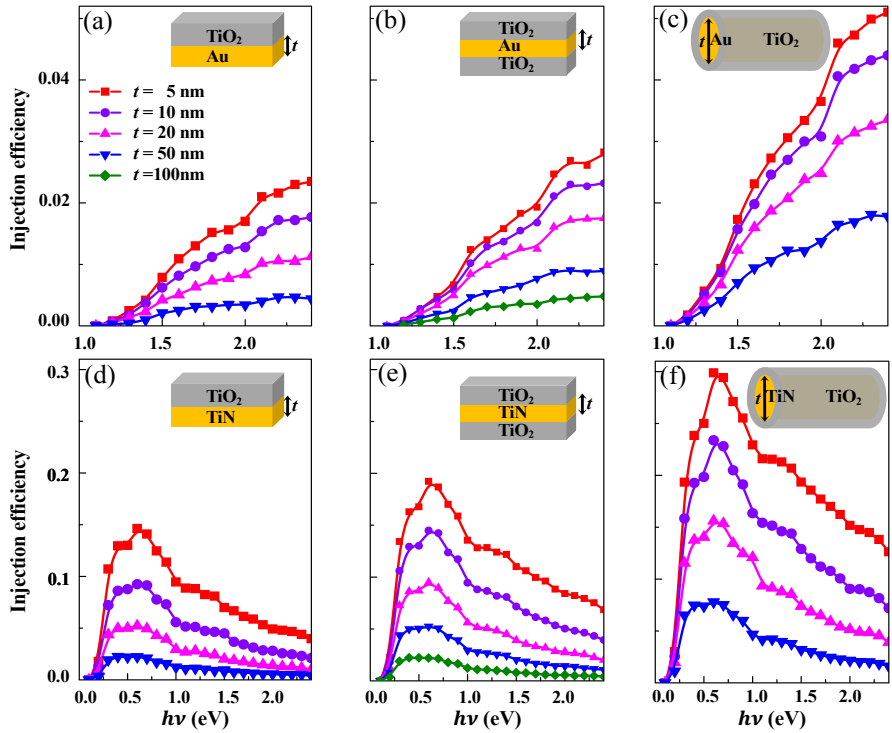

Figure 6. Injection efficiency of hot electrons as a function of incident photon energy $h\nu$ in Au/TiO$_2$ and TiN/TiO$_2$ junctions. The insets depict the structures of the junctions: planar bilayer, planar sandwich trilayer, and core/shell cylinder. $t$ denotes the thickness or diameter of Au and TiN.